# Model-based upscaling of vanadium redox flow battery systems: engineering challenges and solutions

Bence Sziffer [a,b,]*, Viktor Józsa [a]

[a] Department of Energy Engineering, Faculty of Mechanical Engineering, Budapest University of Technology and Economics, Műegyetem rkp. 3., H-1111 Budapest, Hungary

[b] MTA-BME Lendület "Momentum" Renewable Energy Systems Research Group, Műegyetem rkp. 3., H-1111, Budapest, Hungary

**Abstract**

Large-scale energy storage has become an inevitable solution for integrating stochastically available renewable energy sources into the electric grid. Vanadium redox flow batteries offer a viable option among other technologies, due to their long lifetime and independently scalable output power and capacity. The electrolyte temperature should be maintained within the range of 5–40 °C for safe operation; therefore, a thermal management system is necessary, which affects battery efficiency. The study presents a detailed, containerized battery model with a hybrid thermal management system that considers thermal radiation and real ambient temperatures. A total of 180 configurations were investigated in 10 cases, ranging from 4 to 400 kW, with 18 different discharging current–cell number ratios in each case, including multistack arrangements. Current-dependent ohmic losses influence the electric efficiency, which increases from a minimum of 68% to 89% in low-current configurations. However, the net system efficiency ranges between 43% and 66% due to the self-consumption of pumps, the inverter, and the thermal management system. In addition to the detailed efficiency analysis, a comprehensive investigation of thermal processes is provided in terms of the current–cell number ratio and output power, which is crucial for designing thermal management systems and sizing batteries.

* Corresponding author. E-mail: sziffer@energia.bme.hu

**Nomenclature**

Latin letters

| Notation | Description | Unit (if relevant) |
|---|---|---|
| $a$ | absorption coefficient | 1 |
| $A$ | area | m² |
| $c$ | concentration | mol/l |
| $c_p$ | isobaric heat capacity | J/(kgK) |
| $D$ | diffusion coefficient | m²/s |
| $E$ | voltage | V |
| $F$ | Faraday constant | 96485 C/mol |
| $G$ | global irradiance | W/m² |
| $\Delta H$ | reaction enthalpy | kJ/mol |
| $H$ | pump head | m |
| $I$ | current | A |
| $l$ | length | m |
| $L$ | distance | m |
| $N$ | number of cells | 1 |
| $\Delta p$ | pressure loss | Pa |
| $P$ | power | W |
| $\dot{Q}$ | heat flow | W |
| $R$ | resistance | Ω |
| $\Delta s$ | entropy change | kJ/(mol·K) |
| $SoC$ | state of charge | 1 |
| $t$ | time | s |
| $T$ | temperature | K |
| $U$ | overall heat transfer coefficient | W/(m²K) |
| $V$ | volume | m³ |
| $\dot{V}$ | volume flow rate | m³/s |
| $z$ | number of electrons transferred | 1 |

Greek letters

| Notation | Description | Unit |
|---|---|---|
| $\delta$ | thickness | m |
| $E$ | energy consumption | kWh |
| $\eta$ | efficiency | % |
| $\rho$ | density | kg/m³ |
| $\varphi$ | view factor | 1 |

Superscript

| Notation | Description |
| --- | --- |
| n | negative |
| OCV | open circuit voltage |
| p | positive |

Subscript

| Notation | Description |
| --- | --- |
| a,i | inner air |
| c | cell |
| cond | conduction |
| conv | convection |
| cont | container |
| e | electrolyte |
| el | electrode |
| ins | insulation |
| inv | inverter |
| m | membrane |
| p | pipe |
| rad | radiation |
| s | stack |
| t | tank |
| u | unit |

Abbreviations

| Notation | Description |
| --- | --- |
| AC | air conditioner |
| COP | coefficient of performance |
| VRFB | vanadium redox flow battery |

# 1. Introduction

The increasing demand for electricity and the environmental pollution associated with fossil fuels require advancements in the energy system [1]. Thoroughly planned domestic renewable energy generation could decrease dependencies and reduce environmental pollution [2]. However, integrating renewable energy sources with fluctuating power generation into the electric grid is challenging, as they are increasingly problematic to balance with fossil generation alone. Energy storage systems, such as pumped hydro storage, offer a suitable solution with high round-trip efficiency and minimal environmental impact [3] to compensate for predictable daily fluctuations. However, large-scale batteries are more tolerant of rapid switching on and off with low response time [4]. The low availability and ethical issues surrounding the mining process of minerals used in Li-ion batteries draw attention to other emerging technologies [5], such as Na-S or vanadium-redox flow batteries (VRFBs). The latter offers advantages in terms of the abundance of cell materials [6], with a lifetime exceeding 20,000 cycles [7], particularly with electrolyte rebalancing [8]. VRFBs have low fire risk due to the near-room-temperature water-based electrolyte; meanwhile, their response time is in the range of milliseconds [9].

The electrolyte is stored in separate tanks, from which the pumps force it into the cells, where reactions occur. Thanks to the unique architecture, the energy capacity and power rating can be designed independently from each other for the desired application. Despite the clearly advantageous properties of VRFBs, further development is necessary, which underscores the need for modeling [10].

Lumped parameter models are suitable for system-level simulations and determination of system efficiency [11], [12]. Since chemical, electrical, fluid dynamics, and thermal processes are highly interconnected, a comprehensive model is typically required. To accurately determine system efficiency, a detailed thermal model is needed to calculate the energy consumption of the thermal management system. Trovò et al. [13] presented an experimentally validated thermal model of a 9 kW/27 kWh VRFB with a current of up to 400 A, which considers entropic heat from main reactions, irreversible heat from crossover side reactions, and hydraulic losses of the pumps. According to their results, the stack temperature can exceed 50 °C at the end of the long-duration high-current discharge period, which declares the necessity of a cooling system to prevent ion precipitation. Multiple cooling possibilities have been investigated and modeled in the scientific literature. Reynard et al. [14] described a concept in which the electrolyte is cooled using a heat exchanger, and the flow battery is combined with a thermally regenerative electrochemical cycle to further improve battery efficiency by 5-9%. Trovò et al. [15] suggested that maintaining a low flow rate continuously or periodically during standby periods can remarkably decrease the stack temperature, meanwhile keeping the self-discharge losses moderate. Industrial-scale VRFBs are often placed in containers that need to be considered in thermal modeling. Shu et al. [16] considered the container in their model of a 30 kW/130 kWh system and defined the inner air with an additional balance equation in the thermal network. The latter contains the transferred heat from the components and the auxiliary equipment. The authors stated that active cooling, utilizing fans, is necessary when the ambient temperature varies between 25 and 45 °C. In a later article, the authors proposed a hybrid active cooling strategy using an air conditioner (AC) and fans for a 30 kW/240 kWh system, which is necessary in a hot summer environment with an insulated container and long operation times [17]. Wang et al. [18] proposed a room temperature model of a 5 kW/60 kWh VRFB with an AC system and a cooling strategy that can turn off the AC during charging, and keep a maximum room temperature at 25 °C during discharging to save up to 48% of energy used by the AC system, compared to a continuous maximum 30 °C room temperature operation.

A few models can be found in the literature regarding multistack configurations. In reality, stack resistances may differ due to production tolerances. Chen et al. [19] demonstrated in their study that the overall module charging capacity can be improved by arranging stacks with similar resistance in the same branch. The authors presented in a later paper [20] that the effects of transport delay can be minimized with proper pipe geometry and flow rate optimization in a multistack VRFB. Later, Chen et al. [21] presented a simplified thermal model for a 250 kW battery with 8 stacks. Their results showed that the temperature difference between the stacks is negligible. The authors also included electrode permeability variation in their research and demonstrated that stacks with similar electrode permeability should be arranged in the same branch to improve power and capacity [22].

A specific desired power output is feasible with different configurations, depending on the ratio of discharging current, cell number, and stack number. The configurations influence system efficiency. Therefore, this paper aims to present a detailed thermal, electrochemical, and hydraulic model significantly further developed from [23], especially the thermal model and the thermal management system. The model is used to investigate system efficiency and thermal management losses as a function of power output, cell number, and discharging current. A lumped parameter model is presented with 14-27 nodes depending on the number of stacks, supplemented with a hybrid active thermal management system which contains openable control dampers, fans, AC, and electrolyte flushing. A total of 180 configurations are investigated in 10 cases, with power output from 4 kW to 400 kW. Each constant power case is investigated with 18 different cell numbers, limiting the maximal discharging current to 50-250 A.

## 2. Materials and methods

In a VRFB system, the stored energy is released via chemical reactions in the cells. The released electrons can be directed to an external electrical circuit to provide electricity, while the anolyte and catholyte are stored in separate tanks. The reaction in the negative half-cell [24]

$$V^{2+} \underset{\text{discharge}}{\overset{\text{charge}}{\rightleftharpoons}} V^{3+} + e^-, \tag{R1}$$

and in the positive half-cell

$$VO_2^+ + e^- + 2H^+ \underset{\text{discharge}}{\overset{\text{charge}}{\rightleftharpoons}} VO^{2+} + H_2O. \tag{R2}$$

The membrane prevents ion cross diffusion between the electrolytes in the half-cells, but allows proton ($H^+$) diffusion to balance the charge between them. Due to the finite membrane efficiency, self-discharge reactions can occur, causing charge imbalance and generating heat [25], [26]. Once the electrolyte has left the cells, it returns to the storage tank at a lower state of charge (*SoC*). This means it can be recirculated again to provide further electric energy until the *SoC* reaches a preset low value. To accurately model VRFBs, a comprehensive and complex model is necessary due to the interconnection of chemical, electrical, thermal, and hydraulic processes. Concentrated parameter modeling is an effective method for describing the battery at the system level. A multi-physics model was built in Matlab/Simulink R2024a, containing interconnected thermal, chemical, electrical, and hydraulic parts. The containerized battery system was divided into several components, e.g., tanks, in- and outflow pipes, individual stacks, inner air, auxiliary equipment, and individual container walls. The latter is crucial to accurately simulate the effects of ambient air temperature and solar radiation. Unlike other models, the components on the positive and negative sides are treated as distinguished, concentrated parameters. The proposed model was established with the assumptions listed in Table 1.:

Table 1. Modeling assumptions and initial condition.

| Type | Value |
|---|---|
| Electrolyte volume | Constant, added expansion tank. |
| Electrolyte mixing model in the tanks | Continuous stirred-tank reactor (CSTR) |
| Gas formation from side reactions | Neglected |
| Shunt current losses | Neglected |
| Cell temperature and ion concentration | Uniform |
| Initial temperature of each node | $T_0 = 23\ °C$ |

*2.1 Thermal modeling*

A detailed thermal model is essential since the harmful thermal precipitation of vanadium ions occurs above 40 °C [27]. The most significant temperature variation occurs in the stack, where multiple processes affect the electrolyte temperature. Firstly, according to the instantaneous volume flow rate, the anolyte and the catholyte flow into the stack at different temperatures. Simultaneously, the convective heat transfer between the stack and the air in the container influences the stack temperature. The contribution of Joule heat depends on the current and the electrical resistance of the stack [28]. Hence, the membrane cannot completely prevent ion cross-contamination; thus, exothermic side-reactions occur. Additionally, reversible entropic heat reduces the stack temperature during charging and increases it during discharging [13]. The balance equation of the stack temperature is described by Eq. (1):

$$\rho_e(T_s) c_{p,e} V_s \frac{dT_s}{dt} = \dot{V}^p \rho_e(T_s) c_{p,e} \cdot (T_{p,in} - T_s) + \dot{V}^n \rho_e(T_s) c_{p,e} \cdot (T_{p,in} - T_s) + U_s A_s (T_{a,i} - T_s) +$$

$$I^2 R_s + N_c \left( D_{V^{2+}} c_{V^{2+}} (-\Delta H_{(R7)}) + D_{V^{3+}} c_{V^{3+}} (-\Delta H_{(R8)}) + D_{VO^{2+}} c_{VO^{2+}} (-\Delta H_{(R4)}) + \right.$$

$$\left. D_{VO_2^+} c_{VO_2^+} (-\Delta H_{(R5)}) \right) \frac{A_{el}}{\delta_m} + \frac{N \cdot I \cdot T_s (\Delta s^n + \Delta s^p)}{zF} - \dot{Q}_{rad,s}, \tag{1}$$

where $\rho_e(T_s)$ is the arithmetic mean of the anolyte and catholyte densities at the stack temperature, assuming an equal volume of anolyte and catholyte in the stack. $T$ is the temperature, $c_p$ is the isobaric heat capacity, $V$ is the volume, $\dot{V}$ is the volume flow rate, $U$ is the heat transfer coefficient, $I$ is the current, $R$ is the electrical resistance, $N_c$ is the cell number, $D$ is the diffusion coefficient, and $c$ is the concentration. $\Delta H$ denotes the reaction enthalpy of the reaction indicated in subscripts, and $A$ and $\delta$ denote the area and the thickness, respectively. The Faraday constant is denoted by $F$, while $z$ is the number of electrons transferred in a reaction, and $\Delta s^n = (s_{VO}{}^{2+} + s_{H2O} - s_{VO2}{}^+)$ and $\Delta s^p = (s_V{}^{3+} - s_V{}^{2+})$ are entropy changes whose values are assumed to be constant [13]. Subscripts e, s, p, in, a,i, c, el, and m denote the electrolyte, the stack, the pipe, the inflow, the inner air in the container, the cell, the electrode, and the membrane, respectively, while superscripts p and n denote the positive and the negative side. Thermal radiation is denoted by $\dot{Q}_{rad,s}$, which covers the radiative heat transfer between the stack and all the remaining components.

Convective and radiative heat transfer and the transported electrolyte influence the temperature variation of the pipes and tanks. Furthermore, the pumping losses further increase the electrolyte temperature. Equations (2)–(4) describe the temperature variation of the inflow pipes, the outflow pipes, and the tank on the positive side of the battery.

$$\rho_e \left( T_{p,in}^p \right) c_{p,e} V_{p,in}^p \frac{dT_{p,in}^p}{dt} = \dot{Q}_{flow,p,in}^p + \dot{Q}_{conv,p,in}^p + \dot{Q}_{pump,p,in}^p - \dot{Q}_{rad,p,in}^p, \tag{2}$$

$$\rho_e(T_{p,out}^p)c_{p,e}V_{p,out}^p \frac{dT_{p,out}^p}{dt} = \dot{Q}_{flow,p,out}^p + \dot{Q}_{conv,p,out}^p - \dot{Q}_{rad,p,out}^p, \tag{3}$$

$$\rho_e(T_t^p)c_{p,e}V_t \frac{dT_t^p}{dt} = \dot{Q}_{flow,t}^p + \dot{Q}_{conv,t}^p - \dot{Q}_{rad,t}^p, \tag{4}$$

where subscript t denotes the tank, and $\dot{Q}_{pump,p,in} = (1-\eta_{pump}) \cdot P_{pump,in}$ is the heat transferred to the electrolyte, and $P_{pump,in} = \eta_{pump,em} \cdot \eta_{pump,fc} \cdot P_{pump}$ is the input power of the pump, including electric motor and frequency converter efficiencies.

The air within the container is treated as one concentrated parameter. The energy balance equation for this node is calculated as:

$$V_{a,i} \cdot \left( c_{p_{a,i}}(T_{a,i}) \cdot T_{a,i} \cdot \frac{\partial \rho_{a,i}}{\partial t} + \rho_{a,i}(T_{a,i}) \cdot T_{a,i} \cdot \frac{\partial c_{p_{a,i}}}{\partial t} + \rho_{a,i}(T_{a,i}) \cdot c_{p_{a,i}}(T_{a,i}) \cdot \frac{\partial T_{a,i}}{\partial t} \right) = \dot{Q}_{conv,cont} + \dot{Q}_{conv,t}^p + \dot{Q}_{conv,t}^n + \dot{Q}_{conv,p,in}^p + \dot{Q}_{conv,p,in}^n + \dot{Q}_{conv,p,out}^p + \dot{Q}_{conv,p,out}^n + N_s \dot{Q}_{conv,s} + \dot{Q}_{inv,a,i} + 2 \cdot N_{pump} \dot{Q}_{pump,a,i}, \tag{5}$$

where $\dot{Q}_{pump,a,i} = (1-\eta_{pump,em} \cdot \eta_{pump,fc}) \cdot P_{pump}$ denotes the heating of one pump due to the losses of the electric motor and the frequency converter, the coefficient of two considers separate positive and negative sides. A single inverter layout is assumed in each configuration; thus, $\dot{Q}_{inv,a,i} = (1-\eta_{inv}) \cdot P_{inv}$ is the inverter loss. It is recommended to model each container wall separately to account for global irradiance and thermal radiation from the components. Moreover, the walls are crucial in radiative heat transfer thanks to their large surface areas. The walls are welded together via rails that conduct heat and have a temperature-balancing role. The convective heat transfer towards the environment and the inner air side also influences the wall temperature. Summarising the processes, the balance equation of the roof is the following:

$$\left(\rho_{cont} c_{p_{cont}} V_{roof} + \rho_{ins} c_{p_{ins}} \delta_{ins} A_{roof,o}\right) \frac{dT_{roof}}{dt} = \dot{Q}_{conv,roof,o,a} + \dot{Q}_{conv,roof,i,a} +$$

$$G_h A_{roof,o} a_{cont,out} - \dot{Q}_{rad,roof} - \dot{Q}_{rad,roof,ambient} - \dot{Q}_{cond,roof,side1} - \dot{Q}_{cond,roof,side2} -$$

$$\dot{Q}_{cond,roof,front1} - \dot{Q}_{cond,roof,front2}, \qquad (6)$$

where $\dot{Q}_{cond,roof,side1} = k_{cont} \cdot l_{cont} \cdot \delta_{cont} \cdot (T_{roof} - T_{side1})/L_{roof,side1}$ and $\delta_{cont}$ is the thickness of the roof sheet plate and $L_{roof,side1}$ is the distance between the roof and the side of the container, e.g., the thickness of the welding and the top side rail. $G$ is the global irradiance, $a$ is the absorption coefficient. Subscript ins denotes the insulation.

The authors have previously emphasized the importance of radiative heat transfer, especially in large-scale systems with large surface areas [23]. This paper presents a more detailed thermal radiation network that considers each component. The simplified 3D model of the containerised system was created in Ansys Fluent 2023 R1 and is shown in Fig. 1. The same arrangement was applied to each power-capacity configuration, e.g., the cylindrical tanks are at the two ends of the container lengthwise, and the pumps, pipes, and stacks are positioned between them.

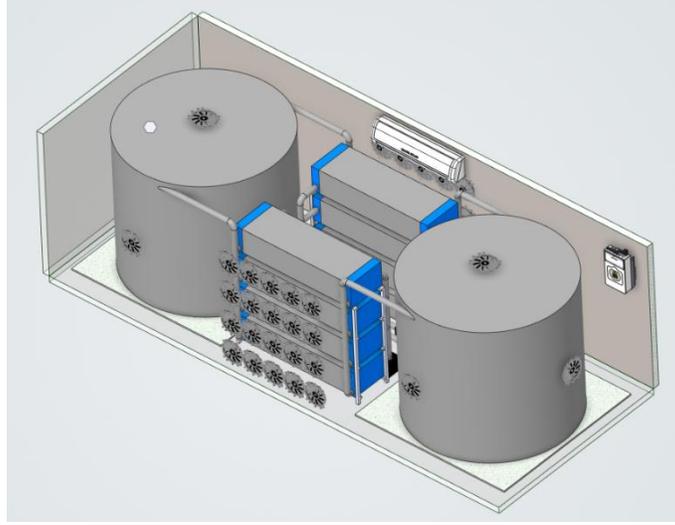

Fig. 1. 3D model of the battery system with 8 stacks.

The view factor matrix was calculated, containing the view factor between each component; an example is presented in Appendix A. Using the view factor matrix, the radiative heat transfer is calculated dynamically as follows:

$$\dot{Q}_{\text{rad},i} = \sum_{j=2}^{13+N_s} \sigma \varepsilon_i \varepsilon_j \varphi_{i,j} A_i \cdot (T_i^4 - T_j^4), \tag{7}$$

where subscript $i$ is any node whose radiative heat transfer is determined, while subscript $j$ denotes the remaining nodes. The view factor between nodes $i$ and $j$ is denoted by $\varphi_{i,j}$. There are 13 nodes in the system, except for the stacks. Hence, the pumps, the inverter, the fans, and the air-conditioner are treated as one node, named auxiliary equipment, whose temperature is assumed to be equal to the inner air temperature.

*2.2 Electrical, chemical, and hydraulic model*

Regarding the electrical model, a *SoC*-based resistance was considered, implemented from Trovò et al. [13]. The cells are assumed to be electrically uniform and connected in series, and the stacks are also connected in series in a multistack arrangement. Therefore, the power of a unit, e.g., the sum of all stack power during discharging, can be determined as follows, where the inner cell resistance decreases the output voltage.

$$P_u = N_s N_c I E_c = N_s N_c I (E_c^{OCV} - R_{c,dch} I) = N_s N_c I \left( E^0 + \frac{\Re T}{zF} \ln \left( \frac{c_{V^{2+}} \cdot c_{VO_2^+} \cdot c_{H^+}^2}{c_{V^{3+}} \cdot c_{VO^{2+}}} \right) - R_{c,dch} I \right), \quad (8)$$

where $E$ denotes the voltage, $\Re = 8.314$ J/(mol·K) is the universal gas constant. Subscript dch denotes discharging, and superscripts OCV and 0 are the open circuit voltage and the standard potential. The chemical part of the multi-physics model determines the actual ion concentration in the stack with balance equations, as detailed in [23].

The hydraulic sub-model calculates the necessary volume flow rate, pump head ($H$), the power consumption of the pumps, and their efficiency using lookup tables based on [29]. Several different flow field designs are employed in VRFBs to minimize the pump work and improve the uniformity of electrolyte distribution [30], like a bifurcate or a conventional interdigitated flow field [31]. The present model assumes the latter; therefore, the total pressure loss of a stack can be determined as the sum of the pressure losses of the inlet and outlet manifolds, plus the cell pressure loss. Moreover, the pressure losses in the pipes also need to be covered by the pump head.

$$H = \frac{\Delta p_s + \Delta p_p}{\rho_e \cdot g}, \quad (9)$$

where $\Delta p$ is the pressure loss and $g$ is the gravitational acceleration. The detailed equations are included in Appendix B.

*2.3 Material properties and input parameters*

Prieto-Díaz et al. [32] measured the density of different vanadium electrolytes at different *SoC* and temperature. As the *SoC* increases, the anolyte density decreases while the catholyte density increases slightly. However, both sides react similarly to temperature changes as the densities decrease with increasing temperature. The opposing behavior as the *SoC* changes implies that the negative and positive sides should be treated as individual concentrated parameters. The electrolyte composition used in the present paper is 2 M vanadium in 4.4 M total sulphates, for which the densities are detailed in Appendix A, while other thermophysical properties are listed in Table 2.

The container walls are usually warmer than the battery components due to the absorbed incident radiation. Therefore, a low-absorption outer surface and a low-emission inner surface are preferable. Considering the year-round operation of the battery, an insulated container may be necessary, depending on the geographical location, using, e.g., metal-clad sandwich panels. The investigation of cold periods and the determination of the proper insulation thickness for the scaled-up battery system is out of the scope of the current paper, thus the previously determined 10 cm container insulation is considered with zincalume cladding, which has 0.27 absorption coefficient in the wavelength range of incident radiation (0.5 μm) and 0.84 emission coefficient in the thermal radiation domain (8.57-9.85 μm) [23], [33]. Dust, precipitation, and other pollutants can alter the absorption and emission coefficients, especially in the solar radiation wavelength range; thus, a 0.5 absorption coefficient and 0.86 emissivity were considered for the container outer walls during the simulations [34]. Slightly polluted white-painted steel and gray-painted polypropylene were considered for the container inner wall and battery components with emissivities of 0.75 and 0.92, respectively. The emission and absorption coefficients were assumed to be the same for the nodes inside the container.

Table 2. Thermophysical properties.

| Notation | Name | Value | Unit | Reference |
|---|---|---|---|---|
| $\rho_e$ | electrolyte density | See Appendix A | [kg/m$^3$] | [35] |
| $c_{p,e}$ | electrolyte heat capacity | 3200 | [J/kgK] | [36] |
| $k_e$ | electrolyte thermal conductivity | 0.33 | [W/mK] | [37] |
| $\mu_e$ | electrolyte dynamic viscosity | 4.928·10$^{-3}$ | [kg/ms] | [36] |
| $\beta_e$ | electrolyte thermal expansion coefficient | 4.488·10$^{-4}$ | [1/K] | [38] |
| $\Phi_{el}$ | graphite electrode porosity | 0.68 | [1] | [36] |
| $k_t = k_p = k_s$ | battery components wall thermal conductivity | 0.22 | [W/mK] | [39] |
| $\varepsilon_t = \varepsilon_p = \varepsilon_s$ | battery components wall emissivity | 0.92 | [1] | [40] |
| $k_{ins}$ | insulation thermal conductivity | 0.025 | [W/mK] | [41] |
| $\rho_{cont}$ | container wall density | 7800 | [kg/m$^3$] | [41] |
| $c_{p,cont}$ | container wall heat capacity | 473 | [J/kgK] | [41] |
| $k_{cont}$ | container wall thermal conductivity | 46 | [W/mK] | [41] |
| $\varepsilon_{cont,o}$ | container outer emissivity | 0.86 | [1] | [33] |
| $a_{cont,o}$ | container outer absorption coefficient | 0.5 | [1] | [33] |
| $\varepsilon_{cont,i}$ | container inner emissivity | 0.75 | [1] | [33] |
| $\varepsilon_{aux,eq}$ | auxiliary equipment emissivity | 0.85 | [1] | [41] |
| $\Delta s^n$ | entropy change in the negative electrolyte | -0.0379 | [kJ/(molK)] | [13] |
| $\Delta s^p$ | entropy change in the positive electrolyte | -0.0884 | [kJ/(molK)] | [13] |

The principal aim of the paper is to evaluate the variation in system efficiency as the battery power and capacity increase from the previously investigated 4 kW system to 400 kW, covering two magnitudes in 10 cases, following the Renard series. The nominal capacity-to-power ratio was 4, common in commercial systems [42], [43], [44]. The key indicators of the 10 setups are summarized in Table 3.

Table 3. Nominal power, nominal capacity, tank volume and height, container length, and width of the investigated battery configurations.

| Case | $P_{nom}$ [kW] | $E_{nom}$ [kWh] | No. of stacks | $V_t$ [l] | $h_t$ [m] | $(l \times wi)_{cont}$ [m] |
|---|---|---|---|---|---|---|
| 1 | 4 | 16 | 1 | 256 | 1 | 2.80×2.35 |
| 2 | 7 | 28 | 1 | 448 | 1.5 | 2.80×2.35 |
| 3 | 11 | 44 | 1 | 704 | 1.5 | 4.34×2.35 |
| 4 | 19 | 76 | 1 | 1216 | 1.8 | 4.34×2.35 |
| 5 | 31 | 124 | 1 | 1984 | 2.2 | 5.89×2.35 |
| 6 | 52 | 208 | 2 | 3328 | 2.2 | 5.89×2.35 |
| 7 | 86 | 344 | 4 | 5504 | 2.2 | 5.89×2.35 |
| 8 | 144 | 576 | 6 | 9216 | 2.2 | 7.30×2.95 |
| 9 | 240 | 960 | 8 | 15360 | 2.2 | 7.30×2.95 |
| 10 | 400 | 1600 | 14 | 25598 | 2.2 | 9.60×4.55 |

Proper ambient temperature and solar irradiance data are crucial for determining thermal processes. Therefore, real weather data with one-minute resolution was used to simulate the ambience precisely. The Baseline Surface Radiation Network provides publicly available ambient temperature and incident radiation data from numerous stations worldwide, from which the Budapest station was selected [45]. The hottest 20-day period of the last available complete year is shown in Fig. 2, from which data from the 4th day to the 14th day were used for the simulations. The first four days were repeated at the beginning to eliminate the effect of thermal inertia differences among the configurations, and later omitted from the calculations.

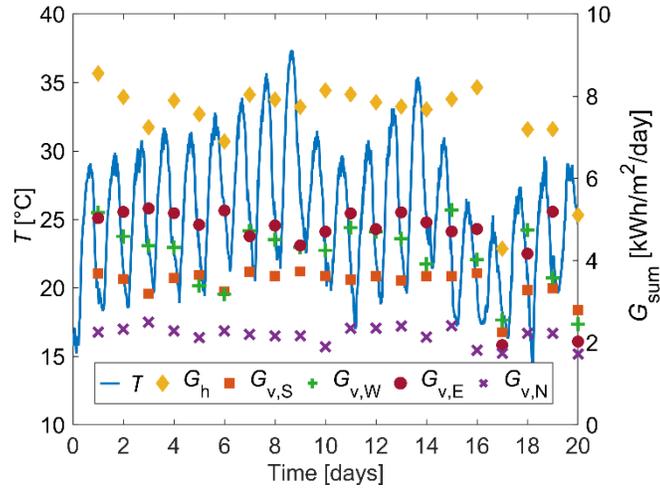

Fig. 2. Ambient temperature and global irradiance in kWh/m$^2$/day by cardinal points. Subscripts h and v mean horizontal and vertical, respectively.

## 3. Results and discussion

### 3.1 Electrical results

Any practical electric power can be delivered with various cell numbers; hence, the cell voltage depends on the discharging current through the ohmic losses of the cell [46]. The higher the discharging current, the higher the ohmic losses, which decrease the cell voltage during discharging. Figure 3 shows the cell voltages obtained from the presented battery model with the same battery configuration and different discharging currents, from the typical current range used in VRFBs [47]. The cell voltage decreases as the discharging current increases, and the cell voltage difference between adjacent curves increases as the *SoC* decreases. Consequently, each battery configuration described in Table 3 was investigated with different cell numbers, with the maximal discharging current limited to the values in Fig. 3. This yields a total of 180 configurations and simulation results, which can be used not only to investigate the battery efficiency with increasing power, but also to provide guidance to find the ideal cell number for typical demands.

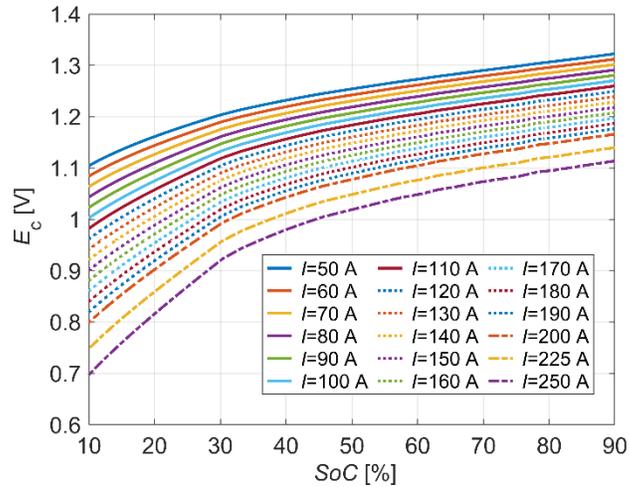
Fig. 3. Cell voltage variation as a function of discharging current.

Constant charging and discharging power were assumed during the simulations for more straightforward comparability. Consequently, the input and output current and the cell voltage are determined according to the actual *SoC* and power demand. Figure 4 shows the daily variation of current and stack voltage, which simulates a battery connected to a photovoltaic (PV) power plant, where it is charged as the PV provides electricity and discharged during peak demand. The presented curves may differ quantitatively but behave identically in different configurations. During charging, between 9:00 and 15:00, the input current initially increases to a higher value due to the low stack voltage. As the stack voltage increases with increasing *SoC*, the input current slightly decreases. An opposite behavior can be observed during discharging between 18:00 and 22:00, due to the decrease in stack voltage resulting from the decreasing SoC; thus, the output current must increase to maintain constant power delivery.

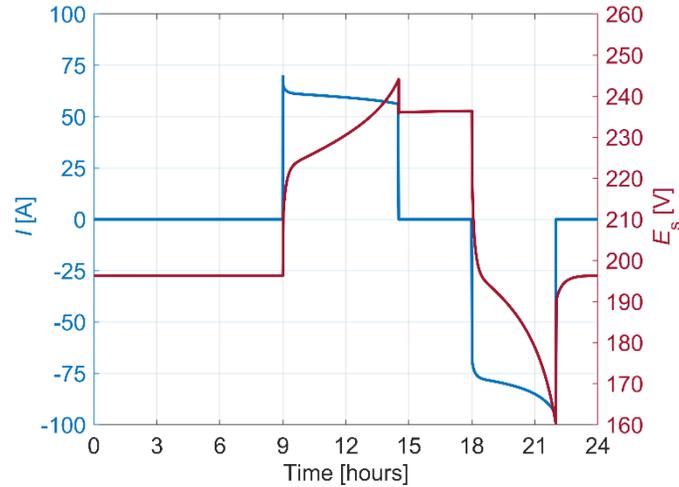
Fig. 4. Daily current and stack voltage variation of Case 4 with 156 cells.

*3.2 Results of single-stack configurations*

The thermal management system features openable control dampers, a single air conditioner, and cooling fans. The fans remove the warm air around the stacks, providing a higher heat transfer coefficient on the stack wall. To achieve a heat transfer coefficient exceeding 10 W/(m$^2$K), the fans are positioned 10 cm before the stack and adjacent to each other, consuming 45 W per unit to operate at the desired working point [48]. Therefore, longer stacks come with increased fan energy consumption. The vertical distance between the stacks was 5 cm in multistack configurations; the narrow gap accelerates the cooling air, thereby increasing the heat transfer coefficient. Four additional fans cool each tank, three on the sides and one on the top. The control logic of thermal management is considered as a 10 W consumption, and favors using the system that delivers the highest actual coefficient of performance (COP) that complies with the temperature limitation, as detailed in [23]. To determine the energy consumption of the AC, the log*p-h* diagram of the assumed R32 cooling fluid was implemented into the model, and complemented by the 65 W fan power of the indoor and outdoor units. During standby periods, the electrolyte is periodically circulated to transport the accumulated heat in the stacks from the self-discharge reactions. One cycle contains a flushing period and a standby period. From the stack volume and the volume flow rate, the electrolyte exchange time in the stack can be determined. The flushing period lasts until one exchange time, while the standby period is twice the exchange time. This method can reduce the energy consumption of the pump by a third compared to continuous electrolyte circulation, while maintaining maximum temperatures 0.3-0.4 °C higher, depending on the electrolyte volume in the stack.

The electric efficiency of the battery accounts for the losses caused by cell resistance, while the system efficiency considers all losses from auxiliary equipment. The former is determined by the ratio of the total discharged energy and the total energy used for charging. The efficiencies of single-stack configurations are shown in Figs. 5a-b, where the crosses denote the simulated configurations, and the presented curves follow a constant power, indicating a power curve relation. It can be seen that the electric efficiency increases as the discharging current decreases; hence, the ohmic losses also decrease. The electric efficiency range spreads from 67.92% ($I = 250$ A, $N = 154$) to 86.18% ($I = 50$ A, $N = 60$). Investigating the variation of electric efficiency at a fixed discharging current shows that dependency on cell number, i.e., output power, is insignificant. The maximum difference of electric efficiency at a fixed discharging current is 1.3% in Fig. 5a. The system efficiency is the ratio of the total discharged energy after the inverter and the total energy used for charging before the inverter, supplemented by the energy usage of circulating pumps, fans, AC, PLC, and flushing. A significant decrement can be observed in Fig. 5b. compared to the electric efficiency; hence, the system efficiency range is between 43.25% ($I = 250$ A, $N = 20$) and 63.63% ($I = 50$ A, $N = 280$). Again, the system efficiency is higher in low current configurations; however, the increment from high current configurations is lower, 10.4% in system efficiency compared to 18.3% in the case of electric efficiency. The difference is consumed by the auxiliary equipment as detailed later. The system efficiency is not uniform across fixed discharging currents, unlike the electric efficiency; hence, it increases as the cell number increases. In the case of $I = 250$ A, the system efficiency difference between the smallest and highest cell number is 7.92% which is reduced to 5.56% at $I = 50$ A. System efficiency is influenced not only by the stack but also by other components and factors, such as container size, pipe length, and tank thermal inertia, which lead to observable differences.

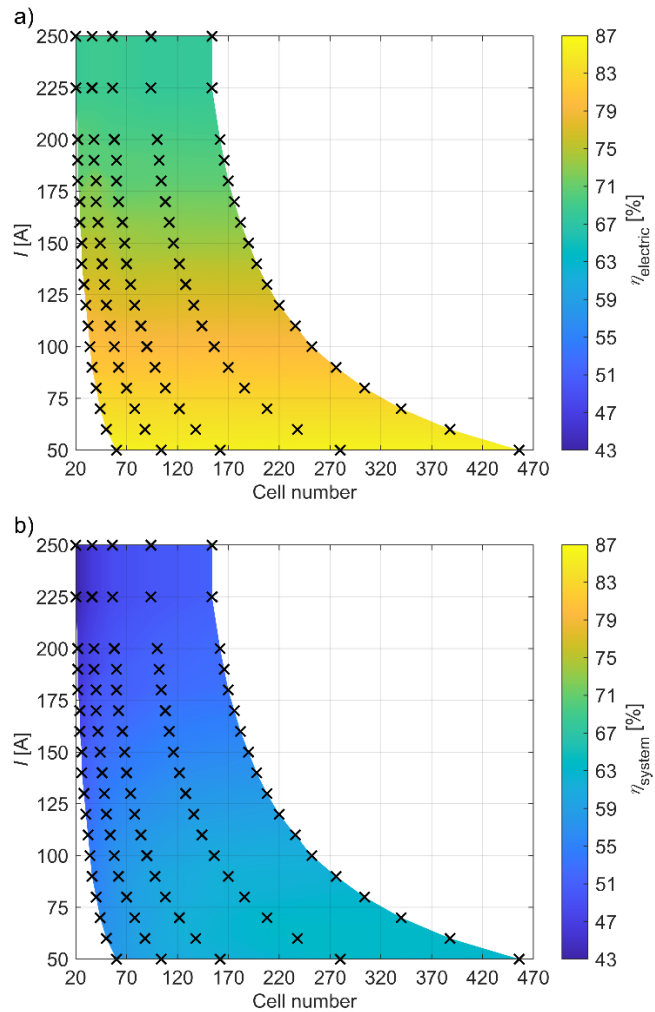

Fig. 5. Efficiency as a function of cell number and discharging current, single stack configurations: a) electric efficiency, b) system efficiency.

The same pump was used in each configuration because it covers all of the occurring pump head and pressure losses at high hydraulic efficiency. Figure 6 presents the efficiency reduction of the pump, showing that the reduction is significantly higher at low power configurations. Several factors explain this. Firstly, the pumps have a minimum power consumption of 64 W even at minimum load. Secondly, the consumption of the pump does not increase to the same extent as the output power of the battery. The total volume flow rate scales almost linearly with battery power due to its strong dependence on $N·I$. However, the total pressure loss of the stack depends primarily on the cell losses, which are characterized by the electrolyte velocity in the cell. The latter is theoretically the same in every configuration at a fixed discharging current due to $\dot{V}_c = \dot{V}_s/N$. The cells are hydraulically connected in parallel, so the total stack pressure loss is determined by the branch with the highest loss. Due to uniform cell losses and the conventional distribution channel arrangement, the branch farthest from the inlet cross-section has the highest losses; consequently, it has the lowest volume flow rate. To achieve uniform power in each cell, the volume flow rate must be adjusted in every branch, resulting in uniform losses throughout the system. The higher the number of cells, the longer the stack, resulting in higher friction caused by manifold pressure loss. However, its contribution to total stack pressure loss is less significant than cell losses. Therefore, the increase in volume flow rate and stack size appears to have a much lesser effect on the final pump power consumption. Pressure losses in the pipes increase as the volume flow rate increases, but their contribution to the total pressure loss is also much lower than the stack pressure losses. For example, in the case of I = 50 A and N = 456, the total volume flow rate and power output are 7.6 and 7.76 times higher than in the case of $I$ = 50 A and $N$ = 60, meanwhile, the maximal pump power consumption is only increased by 1.25 times. It is worth noting that a more optimized, ideal pump could be chosen for every power output, which may decrease the differences, but would not completely eliminate them. The inverter is responsible for a notable power loss, as it reduces the usable electric AC power during discharge and increases the AC power input necessary during charge, which can cover the needed DC power to charge the battery. The reduction is in the range of 11–12%. Hence, the ratio of inverter input power and inverter maximum power determines the inverter efficiency [29]; the inverter maximum power is selected to operate around its maximum efficiency point during charging. Due to the constant charging and discharging power, the power reduction of the inverter remains identical among the constant power configurations.

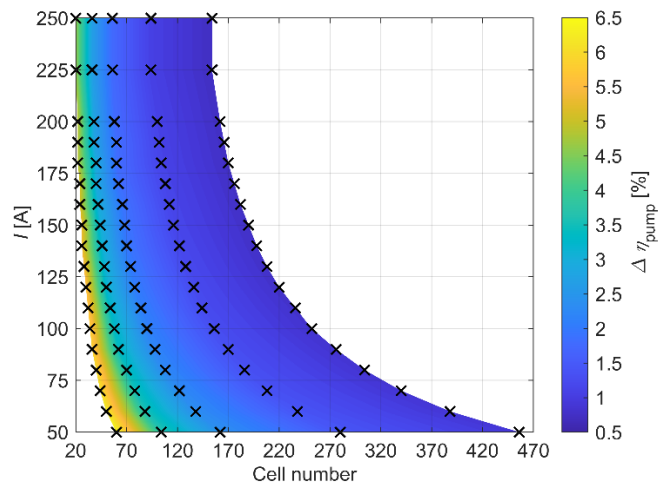

Fig. 6. Efficiency reduction of pumps in single-stack configurations.

Electrolyte temperature is controlled by the fans, AC, and electrolyte circulation. Figures 7a-c illustrate the efficiency reduction resulting from these auxiliary systems. The most significant reduction is due to the AC; hence, its COP is the lowest and has an extended operation time. The contour plots of the fan and AC system show a similar pattern. In high and medium current configurations, the reduction becomes lower at fixed discharging current as the cell number and power output increase. Meanwhile, opposing behaviour can be observed in low current configurations. Interestingly, at a constant output power, the lowest reduction is observed in the medium current configurations. This can be explained by the behaviour of the thermal processes in the stack, which is discussed later in the Section. Flush losses consider the consumption of the pump during electrolyte circulation and the *SoC* loss during this period. The latter has a much smaller contribution to flush losses due to the low diffusion coefficient of vanadium ions across the membrane. The minimum volume flow rate threshold of the circulation pumps causes higher flush losses at smaller power output configurations. A minor advantage can be achieved by selecting a specific pump or adding a low-flow rate pump on a bypass line, which operates only during standby periods. The duration of electrolyte circulation over the 10 days is short, indicating that the overall potential gain is practically negligible. However, a COP greater than 130 can be achieved temporarily through flushing if the flushing flow rate is set to 0.5 l/s and the temperature difference between the stack and the inflow pipe is approximately 4 °C. The lowest modeled COP during flushing is around 3. Figure 7d shows the combined efficiency reduction due to temperature control, indicating that the losses decrease with increasing output power. The lowest efficiency reduction due to thermal management is achieved with the medium current configurations, where the discharging current is 70, 90, 90, 110, and 110 A, respectively, for Cases 1–5. The reason is that the more ideal balance of the current-dependent Joule heat, the current and cell number-dependent entropic heat, and the cell number-dependent heat from self-discharge reactions.

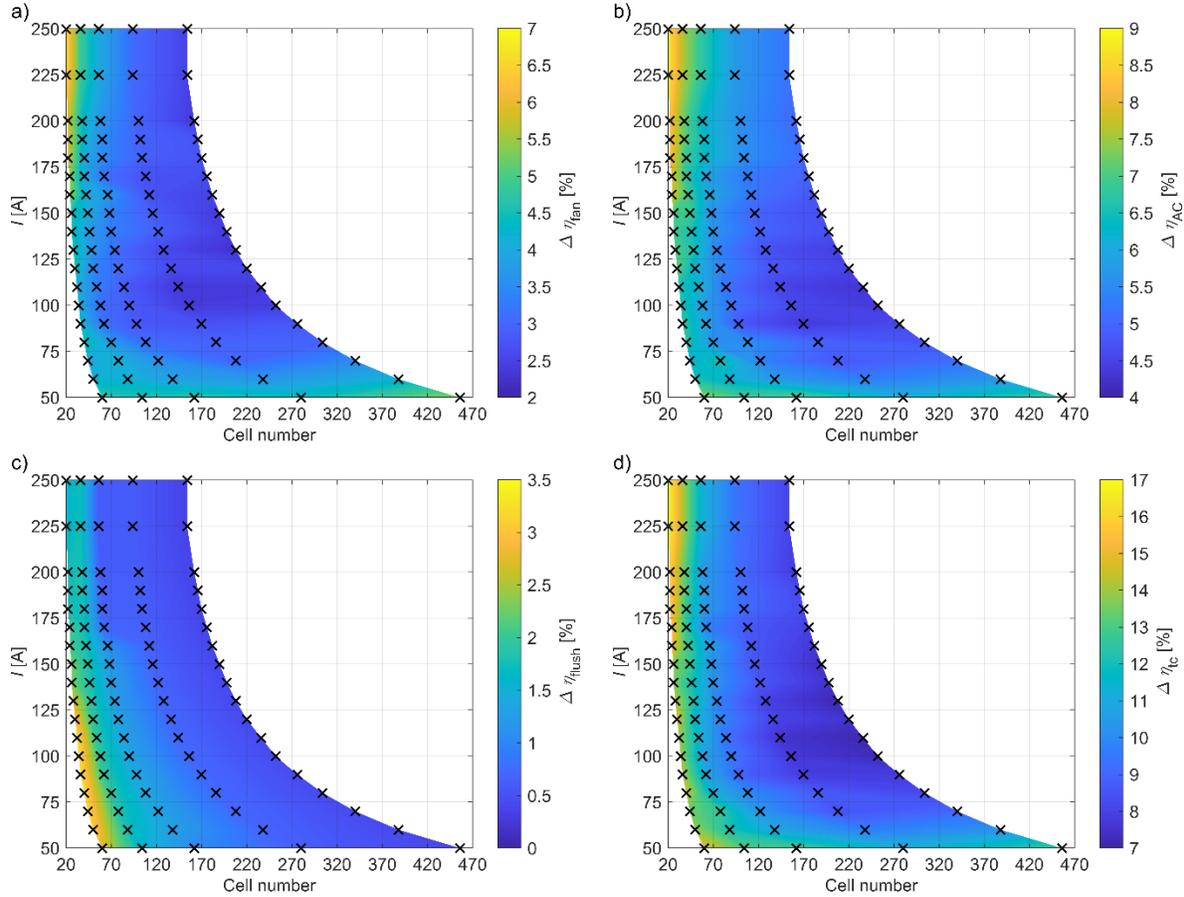

Fig. 7. Efficiency reduction of thermal control in single stack configurations: a) fans, b) air conditioning, c) flush, d) combined temperature control losses.

### 3.3 Results of multistack configurations

The presented methodology was applied to the results of the multistack configurations, which are shown in Figs. 8a-b. The electric efficiency range is between 68.8% ($I = 250$ A, $N_{total} = 1972$) and 88.8% ($I = 50$ A, $N_{total} = 766$); meanwhile, the system efficiency is between 53.1% ($I = 250$ A, $N_{total} = 1972$) and 65.8% ($I = 50$ A, $N_{total} = 650$). The configuration with the highest cell number, Case 5, exhibits the highest system efficiency among single-stack configurations. The multistack configurations are assembled so that the cell number in each stack is close to that of Case 5. There may be minor differences since identical stacks are assumed and only a whole number of stacks is possible.

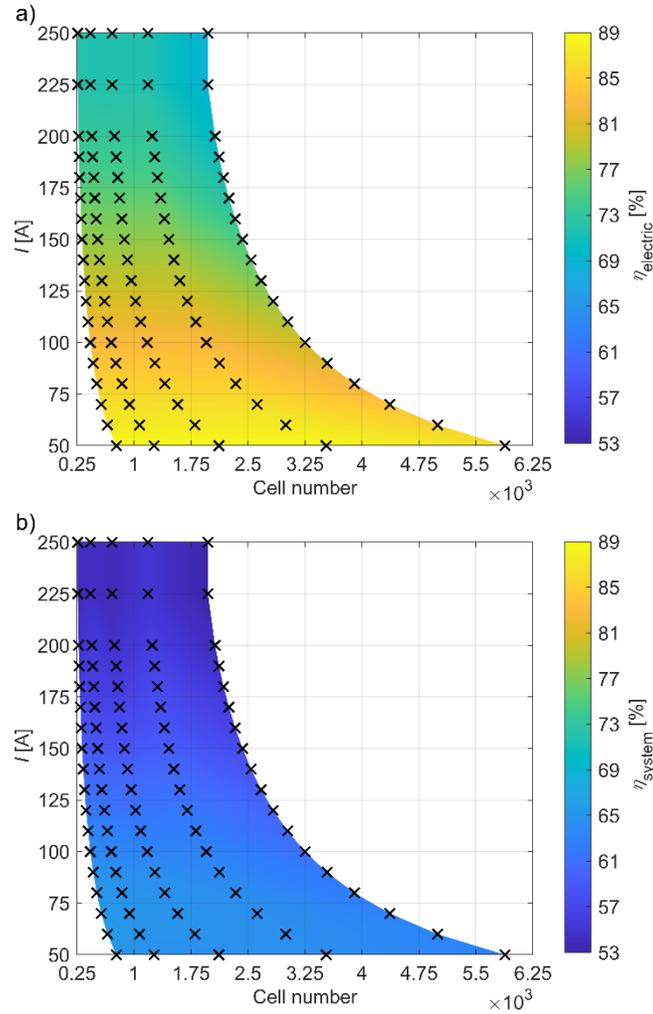
Fig. 8. Efficiency as a function of cell number and discharging current in multistack configurations: a) electric efficiency, b) system efficiency.

In multistack configurations, one pump feeds one stack, which is a safe approach to electricity supply security in the case of a pump malfunction, but may increase investment costs. As shown in Fig. 9, the pumps reduced the total efficiency by 0.96-2.32%. As the maximum current decreases, the energy consumption of the pumps decreases slightly due to the decrease in volume flow rate. As mentioned, the volume flow rate depends on $N·I$, which is higher in high-current configurations.

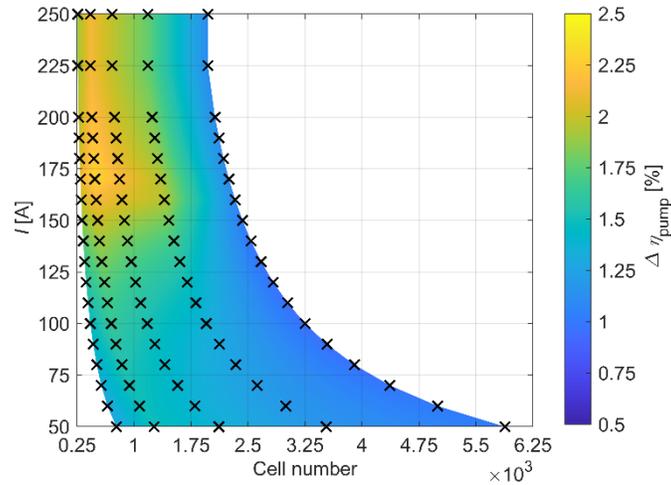
Fig. 9. Efficiency reduction of pumps in multistack configurations.

The efficiency reduction of the thermal management system is shown in Figs. 10a-b, decomposed into the fans and the AC. Both reduce the efficiency to a similar amount as the single-stack Case 5 configuration, with a few exceptions. In the 8- and 14-stack configurations, the power of the tank cooling fans had to be increased from 19 W to 60 W to meet the necessary airflow and heat transfer coefficient requirements; thus, the efficiency reduction is slightly increased. The electrolyte flushing reduces efficiency by only 0–1%. The main conclusions regarding the thermal management system of multistack assemblies indicate that the lowest reduction is observed with medium current configurations in Cases 6–8. Meanwhile, in the case of Cases 9 and 10, the lowest reduction falls within the high current region. However, the difference between high and medium current configurations is below 1% in the latter cases. The thermal management system reduces efficiency by at least 6% and by a maximum of 14.2%, depending on the configuration. This is lower than the single-stack configurations, where the reduction ranged from 7.2% to 17%. The highest reduction is observed in the low current region; however, the difference between the highest and lowest reductions is approximately 6-8%, which is offset by changes in electric efficiency. Since the inverter loss remains constant compared to single-stack configurations, it can be concluded that the highest system efficiency can be achieved with low current and high cell number configurations, regardless of power output and the number of stacks.

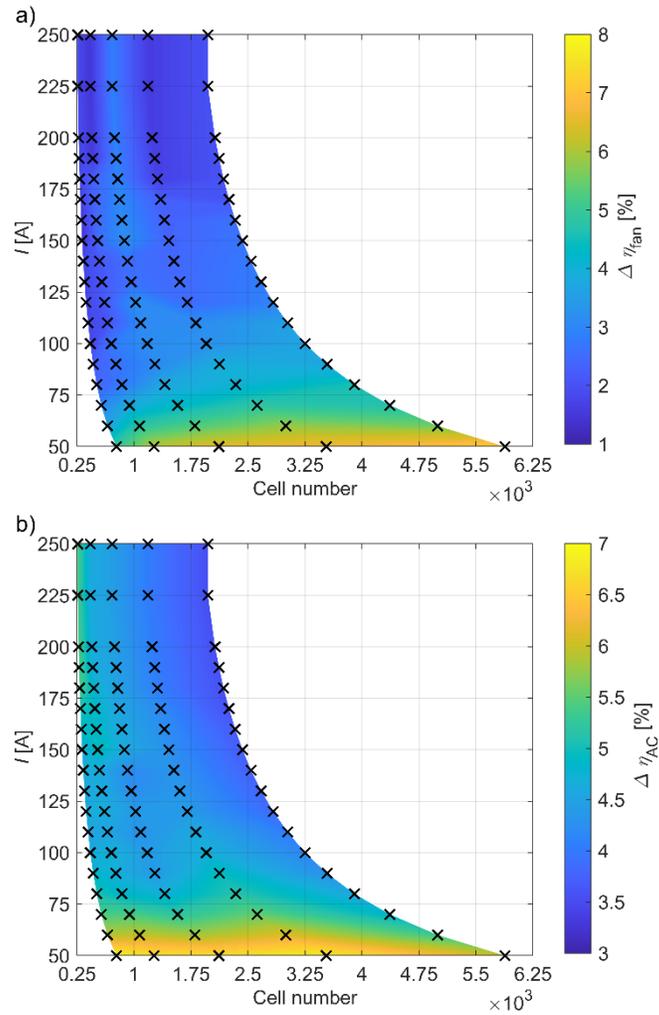

Fig. 10. Efficiency reduction of thermal control in multistack configurations: a) fans, b) air conditioning.

Despite their evident advantages, realizing stacks with high cell numbers may face challenges. Preventing electrolyte leakage in physically large stacks can be difficult, and higher costs and a larger required area can also be limiting factors. Low-current stacks have the advantage that the current can be easily increased, making these stacks more flexible in the event of a cell malfunction. To create an efficient thermal management system, it is essential to investigate the thermal processes in consideration of cell number and current.

*3.4 Investigation of thermal processes*

Several processes influence the stack temperature variation, as described earlier, which determine the operation time and energy consumption of the auxiliary equipment, thereby affecting the overall system efficiency. Figures 11a-d show the amount of heat generated or removed from the processes through a day in the case of the lowest and highest power output single-stack configurations with minimal and maximal cell numbers. The configurations shown in Figs. 11a-b and 11c-d have the same output power. For easier understanding, the moving average of the values is plotted during the standby periods to eliminate sudden changes caused by the periodic electrolyte circulation. During the night, the battery is in standby mode with minimal *SoC*; thus, the pumps are switched off, and the Joule heat is zero.

The stack warmed up during the discharging period the day before, but the air and container walls cooled down during the night, and the thermal management system was operating. Consequently, the stack temperature decreased via thermal radiation and convection. Together with the periodic electrolyte circulation, the generated heat from the self-discharge reactions can be compensated, as indicated by the negative values of $\sum \dot{Q}$. Without electrolyte circulation, the exothermic self-discharge reactions would increase the electrolyte temperature because thermal radiation and heat convection can not withdraw sufficient heat from the stack. In high cell number stack configurations, regardless of the power output, the specific heat from self-discharge reactions is between 2.8 and 3 W/liter; meanwhile, it is around 2 in low cell number configurations. If the circulation is switched off, the geometry of the structure can cause a noteworthy temperature difference among the stacks during the standby periods. As shown in Fig. 11a, flushing was unnecessary; therefore, thermal radiation became the dominant cooling process for the stack. Within a configuration, the smaller the stack, the more sensitive it is to placement and spatial orientation in terms of temperature. Considering the 14-stack configuration with a total of 1,972 cells, the maximum temperature difference among the stacks is 5.7 °C at the end of the longer standby period in the morning and 2.4 °C at the end of the shorter standby period in the evening. Meanwhile, these values are 3 °C and 1.1 °C, with a total of 5,880 cells. However, electrolyte circulation decreases the temperature differences among the stacks; hence, the amount of heat generated by the freshly pumped-in electrolyte is significantly larger than the thermal radiation heat loss. The roof absorbs most solar radiation; therefore, the top stack is the hottest. Consequently, it is recommended to put the stacks on the ground.

During charging, thermal radiation and convection are inferior processes in heat generation; thus, the temperature of the stacks becomes homogeneous. At the beginning of the charging, a short transient heat withdrawal may be observed regardless of the cell number. The reactions are practically instantaneous, while electrolyte pumping takes time. Therefore, the heat from the enthalpy of the endothermic charging reactions is initially greater than the sum of the heat from all other processes combined. Initially, the tank temperature is larger than the stack temperature due to the higher thermal inertia. The stack temperature remains almost constant after the initial decrease; thus, the tanks cool down during charging. As the temperature difference between the tanks and stacks decreases, accompanied by a slight reduction in Joule heat, the significance of the pumped-in electrolyte in heat generation decreases slightly. The cell voltage increases as the $SoC$ increases; thus, the charging current should be reduced to maintain the constant power charging. Moreover, the cell resistance also decreases with increasing $SoC$, which further reduces the Joule heat. Due to the incident solar radiation, the temperature of the container walls increases. Still, the heat absorbed by the stacks from thermal radiation is lower than that from other thermal processes. The heat from exothermic self-discharge reactions also increases as the $SoC$ increases, because the reactions with the two highest enthalpies are related to high $V^{2+}$ and $VO_2^+$ concentrations, which are high if the battery is charged [23].

Figure 11c illustrates an example between 12 and 15 hours of the AC and cooling fan operation. Between the charging and discharging periods, the stack temperature can be controlled using the proposed thermal management system, despite the significant heat generation from self-discharge reactions, which results in a temperature increment of ~3 °C, regardless of the cell number. However, this increment could be reduced by increasing the flushing flow rate or cycle time.

At the start of the discharging period, the heat generation suddenly increases due to the instantaneous exothermic reactions and Joule heat. Again, a short time is necessary for the freshly pumped-in electrolyte to flow through the stack. As the $SoC$ decreases, the Joule heat increases due to the increasing absolute value of the current, compensating for the decreasing stack voltage to maintain the constant discharging power. The increasing absolute value of the current slightly increases the heat generated by the main reaction entropy as well. Moreover, the volume flow rate also needs to be increased as the $SoC$ decreases to maintain the constant output power. The volume flow rate also depends on the discharging current, whose absolute value increases during discharging, thereby further enhancing the cooling effect of the pumped-in electrolyte. The more significant decrement in cell voltage in high current configurations results in more pronounced changes in discharging current, thus in Joule heat, volume flow rate, and heat from main reaction entropy, compared to low current configurations. Considering the discharging period at night, when the container walls are cooler than the stack, thermal radiation enhances heat rejection.

A flow factor of 2 was used during charging and 5 during discharging to reduce the warming effect of pumped-in electrolyte during charging and to exploit the colder electrolyte in the tanks for cooling during discharging. The resulting volume flow rate range is hence wide for a single unit, which limits the selection of suitable devices; however, a two-stage pump system can also be a solution. The volume flow rate ranges from 0.6 to 30 m³/h, and the pump head varies between 0.2 and 15 m in the investigated configurations. It is clear that the pumped-in electrolyte has a dominant role in the temperature variation of the stack; therefore, a dynamically modulated flow factor could be an efficient solution for thermal management.

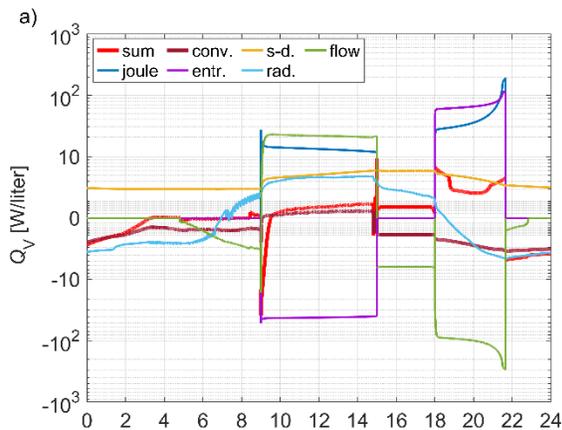
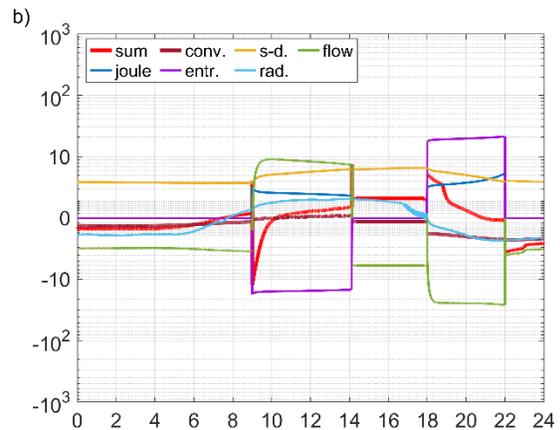
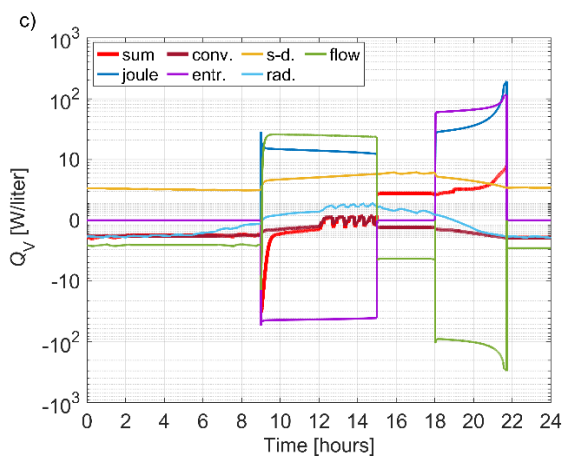
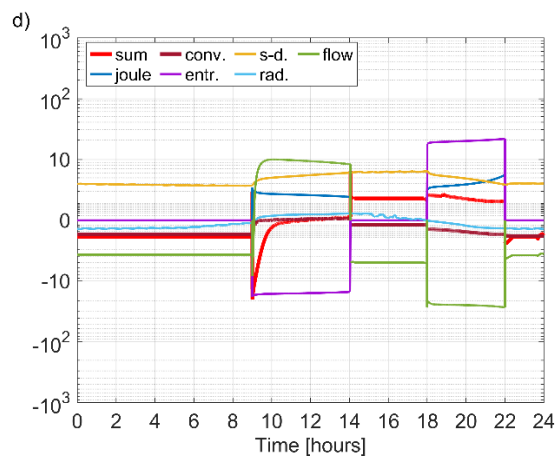

Fig. 11. Daily heat variation in the stack, a) Case 1: $N = 20$, $I = 250$ A, b) Case 1: $N = 60$, $I = 50$ A, c) Case 5: $N = 154$, $I = 250$ A, d) Case 5: $N = 456$, $I = 50$ A.

## 4. Conclusions

This study presents a concentrated parameter model of a containerized vanadium redox flow battery, supplemented by the detailed models of the auxiliary equipment and thermal management system. A total of 180 different configurations were simulated to investigate system efficiency as the battery output power was scaled up from 4 to 400 kW in 10 cases, and the discharging current varied from 50 to 250 A in 18 steps. The same one-pump-per-stack layout, spatial arrangement, cell size, charge-discharge profile, ambient conditions, and power-to-capacity ratio were used during upscaling to improve comparability. Based on the results, the following conclusions were drawn.

At a fixed power output, the lower the discharging current, the higher the electrical efficiency due to the decrease in ohmic losses. The electric efficiency range is between 68% and 89%, with the output power causing only minor differences within this range.

The system efficiency range is between 43 and 66%, and it is higher in low-current configurations. However, it also depends on the power output of the stack; the higher the stack power output, the higher the system efficiency. The maximum achieved system efficiency is 58–60% in low-power output configurations and 64–66% in high-power output configurations.

The most significant losses are caused by the thermal management system, ranging from 7.5% to 17% in low power output configurations, especially at minimal and maximal discharging currents, followed by the inverter at 11–12%, and then the pumping losses at 2–6.5%. However, the thermal management losses are decreased to 6–14.2% in high power output configurations, and the pumping losses are also reduced to 1–2%.

The presented simulation results provide a large amount of data regarding the efficiency of the vanadium redox flow battery during upscaling. The study highlights the key areas for future work to reduce the consumption of thermal management systems, like the exploitation of dynamic flow factor modulation, the two-stage pumping system for electrolyte circulation to utilize its high COP even in low power stacks. Another research area could be the potential exploitation of the generated waste heat through a thermally regenerative electrochemical cycle to produce electricity.

**Funding**

The research reported in this paper was supported by the 2021-2.1.1-EK-00001 project within the framework of the 2021-2.1.1-EK program, the DKÖP-25-1-BME-84, OTKA-FK 137758, and NKKP ADVANCED 150696 funded by the National Research, Development and Innovation Fund of Hungary, and the János Bolyai Research Scholarship of the Hungarian Academy of Sciences.

**Appendix A**

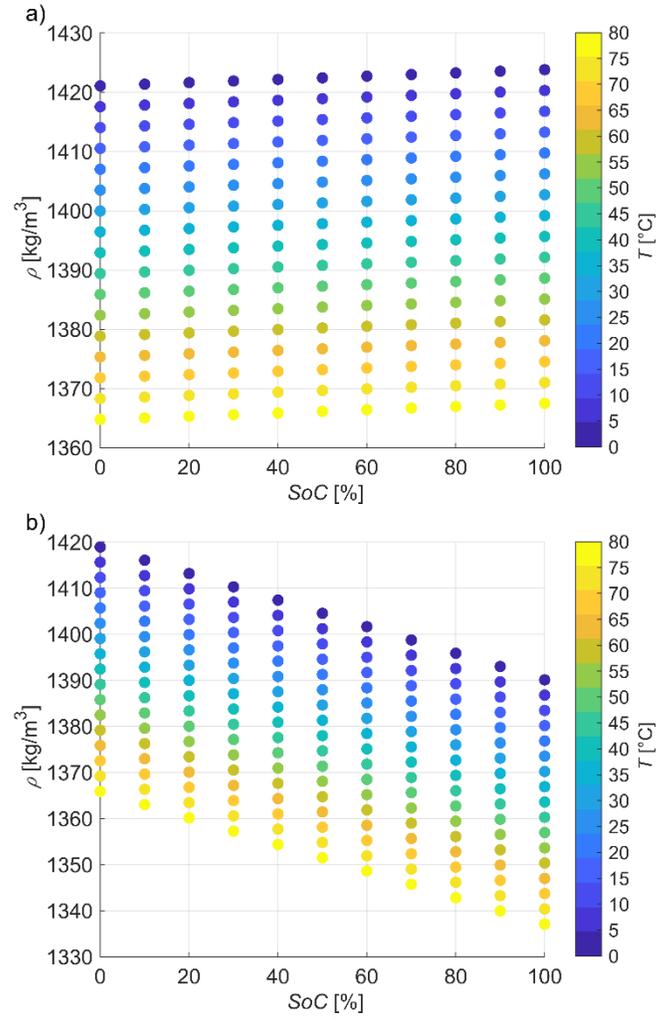

Fig. A1. Electrolyte density as a function of temperature and *SoC*, a) catholyte, b) anolyte.

Table A.1. View factor matrix of a containerized VRFB system with four stacks.

| | Aux.eq | Floor | Front1 | Front2 | Pipe,in$^n$ | Pipe,in$^p$ | Pipe,out$^n$ | Pipe,out$^p$ | Roof | Side1 | Side2 | Stack1 | Stack2 | Stack3 | Stack4 | Tank$^n$ | Tank$^p$ |
|---|---|---|---|---|---|---|---|---|---|---|---|---|---|---|---|---|---|
| Aux. eq | 0.048 | 0.088 | 0.025 | 0.051 | 0.013 | 0.014 | 0.009 | 0.010 | 0.087 | 0.177 | 0.089 | 0.047 | 0.043 | 0.048 | 0.037 | 0.072 | 0.125 |
| Floor | 0.107 | 0.000 | 0.023 | 0.022 | 0.023 | 0.029 | 0.005 | 0.005 | 0.093 | 0.147 | 0.195 | 0.127 | 0.020 | 0.011 | 0.006 | 0.091 | 0.093 |
| Front1 | 0.035 | 0.033 | 0.000 | 0.002 | 0.000 | 0.000 | 0.001 | 0.000 | 0.083 | 0.124 | 0.124 | 0.000 | 0.000 | 0.000 | 0.000 | 0.600 | 0.002 |
| Front2 | 0.070 | 0.032 | 0.001 | 0.000 | 0.000 | 0.000 | 0.000 | 0.001 | 0.081 | 0.098 | 0.126 | 0.000 | 0.000 | 0.000 | 0.000 | 0.002 | 0.595 |
| Pipe,in$^n$ | 0.115 | 0.193 | 0.000 | 0.000 | 0.024 | 0.033 | 0.000 | 0.001 | 0.066 | 0.032 | 0.158 | 0.080 | 0.052 | 0.044 | 0.020 | 0.053 | 0.121 |
| Pipe,in$^p$ | 0.096 | 0.179 | 0.000 | 0.000 | 0.025 | 0.032 | 0.001 | 0.000 | 0.095 | 0.028 | 0.196 | 0.037 | 0.044 | 0.039 | 0.024 | 0.139 | 0.057 |
| Pipe,out$^n$ | 0.089 | 0.045 | 0.006 | 0.002 | 0.000 | 0.001 | 0.016 | 0.005 | 0.136 | 0.334 | 0.012 | 0.043 | 0.050 | 0.053 | 0.057 | 0.137 | 0.002 |
| Pipe,out$^p$ | 0.099 | 0.050 | 0.003 | 0.004 | 0.001 | 0.000 | 0.005 | 0.017 | 0.139 | 0.339 | 0.022 | 0.043 | 0.051 | 0.054 | 0.059 | 0.002 | 0.123 |
| Roof | 0.050 | 0.048 | 0.034 | 0.034 | 0.004 | 0.008 | 0.008 | 0.008 | 0.000 | 0.103 | 0.163 | 0.003 | 0.005 | 0.010 | 0.067 | 0.228 | 0.229 |
| Side1 | 0.105 | 0.079 | 0.052 | 0.041 | 0.002 | 0.002 | 0.021 | 0.021 | 0.109 | 0.000 | 0.031 | 0.036 | 0.042 | 0.037 | 0.033 | 0.209 | 0.181 |
| Side2 | 0.050 | 0.098 | 0.049 | 0.049 | 0.010 | 0.016 | 0.001 | 0.001 | 0.161 | 0.030 | 0.000 | 0.016 | 0.022 | 0.023 | 0.022 | 0.226 | 0.228 |
| Stack1 | 0.127 | 0.299 | 0.000 | 0.000 | 0.023 | 0.014 | 0.011 | 0.011 | 0.014 | 0.161 | 0.080 | 0.000 | 0.178 | 0.000 | 0.000 | 0.042 | 0.041 |
| Stack2 | 0.117 | 0.048 | 0.001 | 0.000 | 0.015 | 0.017 | 0.013 | 0.013 | 0.024 | 0.185 | 0.107 | 0.176 | 0.000 | 0.178 | 0.000 | 0.054 | 0.053 |
| Stack3 | 0.130 | 0.026 | 0.001 | 0.000 | 0.013 | 0.015 | 0.014 | 0.014 | 0.046 | 0.169 | 0.110 | 0.000 | 0.177 | 0.000 | 0.178 | 0.056 | 0.054 |
| Stack4 | 0.100 | 0.014 | 0.000 | 0.000 | 0.006 | 0.009 | 0.015 | 0.015 | 0.312 | 0.146 | 0.102 | 0.000 | 0.000 | 0.176 | 0.000 | 0.050 | 0.051 |
| Tank$^n$ | 0.039 | 0.044 | 0.228 | 0.001 | 0.003 | 0.011 | 0.007 | 0.000 | 0.213 | 0.187 | 0.216 | 0.008 | 0.011 | 0.011 | 0.010 | 0.000 | 0.011 |
| Tank$^p$ | 0.067 | 0.045 | 0.001 | 0.226 | 0.007 | 0.004 | 0.000 | 0.006 | 0.215 | 0.162 | 0.217 | 0.008 | 0.011 | 0.011 | 0.010 | 0.011 | 0.000 |

As an example, the view factor between side 1 and the roof for radiation from side 1 to the roof is 0.109 (tenth row, ninth column).

The 10 power cases based on the Renard series can be determined according to the following:

$$P_n = P_1 \cdot r^n, \tag{A.1}$$

where $P_1 = 4$ kW, $n = 10$ is the number of cases, and $r$ is the common ratio:

$$r = \left(\frac{P_{10}}{P_1}\right)^{\frac{1}{n-1}} = \left(\frac{400}{4}\right)^{\frac{1}{10-1}} = 1.668, \tag{A.2}$$

**Appendix B**

The total pressure loss of a stack is the sum of manifold pressure losses and the pressure loss of the cells with an interdigitated flow field:

$$\Delta p_s = \left(\Delta p_{man,i} + \Delta p_c + \Delta p_{man,o}\right), \tag{B.1}$$

where subscript man denotes the manifold. The manifold pressure loss consists of losses in the straight sections and the T-junctions:

$$\Delta p_{man} = \Delta p_{man,str} + \Delta p_{man,T,str} + \Delta p_{man,T,br}, \tag{B.2}$$

where

$$\Delta p_{\text{man,str}} = \frac{\rho w^2 (\delta_c - \delta_{\text{ch}}) \xi_{\text{str}}}{2 d_{\text{man}}}, \xi_{\text{str}} = \frac{64}{\text{Re}_{\text{man}}}, \text{Re}_{\text{man}} = \frac{w d_{\text{man}}}{\nu}, \tag{B.3}$$

$$\Delta p_{\text{man,T,str}} = \xi_{\text{T,str}} \rho \frac{w^2}{2}, \xi_{\text{T,str}} = 0.2, w = \frac{\dot{V}_c}{\frac{d_{\text{man}}^2 \pi}{4}}, \tag{B.4}$$

$$\Delta p_{\text{man,T,br}} = \xi_{\text{T,br}} \rho \frac{w^2}{2}, \xi_{\text{T,br}} = 1, \tag{B.5}$$

where subscript str, T, and br denote straight, T-junction, and branch. The pressure loss of an interdigitated flow can be calculated using a hydrodynamic network [49]. The pressure loss of the cell with the maximum pressure loss is considered the cell pressure loss; hence, the cells are connected hydraulically in parallel.

$$\Delta p_c = R_{\text{iff}} \dot{V}_c \rho_e(T_s), \tag{B.6}$$

$$R_{\text{iff}} = \frac{\mu \text{Re}_f w i_c (\delta_{\text{ch}} + h_{\text{ch}})^2}{2 \rho \delta_{\text{ch}}^3 h_{\text{ch}}^3} + \frac{\mu \text{Re}_f w i_c (\delta_{\text{ch}} + h_{\text{ch}})^2}{2 \rho \delta_{\text{ch}}^3 h_{\text{ch}}^3 n} + \frac{\mu (\delta_{\text{rib}} + \delta_{\text{el}} + 0.5 \delta_{\text{ch}})}{K \rho ((2n-1) L_{\text{ch}} + w i_c) \delta_{\text{el}}}, \tag{B.7}$$

$$\text{Re}_f = 13.84 + 10.38 e^{\frac{-3.4}{b}}, \tag{B.8}$$

$$b = \frac{\delta_{\text{ch}}}{h_{\text{ch}}}, \tag{B.9}$$

$$n = \frac{w i_c}{\delta_{\text{ch}} + \delta_{\text{rib}}}, \tag{B.10}$$

$$K = \frac{d_f^2 \phi^3}{C(1-\phi)^2}, \tag{B.11}$$

where $wi_c$ denotes the cell width, $d_f$ is the diameter of the graphite fiber in the electrode, $\Phi$ is the porosity, and $C$ is the Carman-Kozeny constant. The dimensions of the assumed channel are the following: $\delta_{\text{ch}} = 2$ mm, $h_{\text{ch}} = 2$ mm, $\delta_{\text{rib}} = 2$ mm. The pressure loss of the pipe system can be determined as the sum of the pressure losses of the pipe elements in the inflow and outflow pipes.

$$\Delta p_\text{p}^\text{p} = 2 \cdot \left[\xi_\text{str} \frac{\rho_\text{e}}{2} w^2 \frac{L_\text{p}}{d_\text{p}} + N_\text{elb}\left(\xi_\text{elb} \frac{\rho_\text{e}}{2} w^2\right)\right], \tag{B.12}$$

where subscript elb denotes the elbows, and $\xi_\text{elb} = 0.3$.

**References**

[1]   B. J. van Ruijven, E. De Cian, and I. Sue Wing, "Amplification of future energy demand growth due to climate change," *Nature Communications 2019 10:1*, vol. 10, no. 1, pp. 1–12, Jun. 2019, doi: 10.1038/s41467-019-10399-3.

[2]   Q. Hassan, S. Algburi, A. Z. Sameen, H. M. Salman, and M. Jaszczur, "A review of hybrid renewable energy systems: Solar and wind-powered solutions: Challenges, opportunities, and policy implications," *Results in Engineering*, vol. 20, p. 101621, Dec. 2023, doi: 10.1016/J.RINENG.2023.101621.

[3]   S. Rehman, L. M. Al-Hadhrami, and M. M. Alam, "Pumped hydro energy storage system: A technological review," *Renewable and Sustainable Energy Reviews*, vol. 44, pp. 586–598, Apr. 2015, doi: 10.1016/J.RSER.2014.12.040.

[4]   Z. Zhu *et al.*, "Rechargeable Batteries for Grid Scale Energy Storage," Nov. 23, 2022, *American Chemical Society*. doi: 10.1021/acs.chemrev.2c00289.

[5]   X. Sun, H. Hao, P. Hartmann, Z. Liu, and F. Zhao, "Supply risks of lithium-ion battery materials: An entire supply chain estimation," *Mater Today Energy*, vol. 14, p. 100347, Dec. 2019, doi: 10.1016/J.MTENER.2019.100347.

[6]   A. Aluko and A. Knight, "A Review on Vanadium Redox Flow Battery Storage Systems for Large-Scale Power Systems Application," *IEEE Access*, vol. 11, pp. 13773–13793, 2023, doi: 10.1109/ACCESS.2023.3243800.

[7]   H. R. Jiang, J. Sun, L. Wei, M. C. Wu, W. Shyy, and T. S. Zhao, "A high power density and long cycle life vanadium redox flow battery," *Energy Storage Mater*, vol. 24, pp. 529–540, Jan. 2020, doi: 10.1016/J.ENSM.2019.07.005.

[8]   N. Poli, M. Schäffer, A. Trovò, J. Noack, M. Guarnieri, and P. Fischer, "Novel electrolyte rebalancing method for vanadium redox flow batteries," *Chemical Engineering Journal*, vol. 405, p. 126583, Feb. 2021, doi: 10.1016/J.CEJ.2020.126583.

[9]  Y. Li, J. Bao, M. Skyllas-Kazacos, M. P. Akter, X. Zhang, and J. Fletcher, "Studies on dynamic responses and impedance of the vanadium redox flow battery," *Appl Energy*, vol. 237, pp. 91–102, Mar. 2019, doi: 10.1016/J.APENERGY.2019.01.015.

[10] Z. Huang, A. Mu, L. Wu, B. Yang, Y. Qian, and J. Wang, "Comprehensive Analysis of Critical Issues in All-Vanadium Redox Flow Battery," *ACS Sustain Chem Eng*, vol. 10, no. 24, pp. 7786–7810, Jun. 2022, doi: 10.1021/ACSSUSCHEMENG.2C01372.

[11] T. Puleston, A. Clemente, R. Costa-Castelló, and M. Serra, "Modelling and Estimation of Vanadium Redox Flow Batteries: A Review," *Batteries 2022, Vol. 8, Page 121*, vol. 8, no. 9, p. 121, Sep. 2022, doi: 10.3390/BATTERIES8090121.

[12] S. Bogdanov *et al.*, "Dynamic modeling of vanadium redox flow batteries: Practical approaches, their applications and limitations," *J Energy Storage*, vol. 57, p. 106191, Jan. 2023, doi: 10.1016/J.EST.2022.106191.

[13] A. Trovò, A. Saccardo, M. Giomo, and M. Guarnieri, "Thermal modeling of industrial-scale vanadium redox flow batteries in high-current operations," *J Power Sources*, vol. 424, pp. 204–214, Jun. 2019, doi: 10.1016/J.JPOWSOUR.2019.03.080.

[14] D. Reynard, C. R. Dennison, A. Battistel, and H. H. Girault, "Efficiency improvement of an all-vanadium redox flow battery by harvesting low-grade heat," *J Power Sources*, vol. 390, pp. 30–37, Jun. 2018, doi: 10.1016/J.JPOWSOUR.2018.03.074.

[15] A. Trovò and M. Guarnieri, "Standby thermal management system for a kW-class vanadium redox flow battery," *Energy Convers Manag*, vol. 226, Dec. 2020, doi: 10.1016/j.enconman.2020.113510.

[16] B. Shu, L. S. Weber, M. Skyllas-Kazacos, J. Bao, and K. Meng, "Thermal Modelling and Simulation Studies of Containerised Vanadium Flow Battery Systems," *Batteries 2023, Vol. 9, Page 196*, vol. 9, no. 4, p. 196, Mar. 2023, doi: 10.3390/BATTERIES9040196.

[17] B. Shu, M. Skyllas-Kazacos, J. Bao, and K. Meng, "Hybrid Cooling-Based Thermal Management of Containerised Vanadium Flow Battery Systems in Photovoltaic Applications," *Processes 2023, Vol. 11, Page 1431*, vol. 11, no. 5, p. 1431, May 2023, doi: 10.3390/PR11051431.

[18] H. Wang, W. L. Soong, S. A. Pourmousavi, X. Zhang, N. Ertugrul, and B. Xiong, "Thermal dynamics assessment of vanadium redox flow batteries and thermal management by active temperature control," *J Power Sources*, vol. 570, Jun. 2023, doi: 10.1016/j.jpowsour.2023.233027.

[19] H. Chen, S. Wang, H. Gao, X. Feng, C. Yan, and A. Tang, "Analysis and optimization of module layout for multi-stack vanadium flow battery module," *J Power Sources*, vol. 427, pp. 154–164, Jul. 2019, doi: 10.1016/J.JPOWSOUR.2019.04.054.

[20] H. Chen, X. Li, H. Gao, J. Liu, C. Yan, and A. Tang, "Numerical modelling and in-depth analysis of multi-stack vanadium flow battery module incorporating transport delay," *Appl Energy*, vol. 247, pp. 13–23, Aug. 2019, doi: 10.1016/J.APENERGY.2019.04.034.

[21] F. Chen, H. Gao, H. Chen, and C. Yan, "Evaluation of thermal behaviors for the multi-stack vanadium flow battery module," *J Energy Storage*, vol. 27, Feb. 2020, doi: 10.1016/J.EST.2019.101081.

[22] H. Chen, M. Cheng, X. Feng, Y. Chen, F. Chen, and J. Xu, "Analysis and optimization for multi-stack vanadium flow battery module incorporating electrode permeability," *J Power Sources*, vol. 515, p. 230606, Dec. 2021, doi: 10.1016/J.JPOWSOUR.2021.230606.

[23] B. Sziffer, M. J. Mayer, and V. Józsa, "Detailed system modeling of a vanadium redox flow battery operating at various geographical locations," *Appl Energy*, vol. 384, p. 125473, Apr. 2025, doi: 10.1016/J.APENERGY.2025.125473.

[24] C. Choi *et al.*, "Understanding the redox reaction mechanism of vanadium electrolytes in all-vanadium redox flow batteries," *J Energy Storage*, vol. 21, pp. 321–327, Feb. 2019, doi: 10.1016/J.EST.2018.11.002.

[25] J. Sun, D. Shi, H. Zhong, X. Li, and H. Zhang, "Investigations on the self-discharge process in vanadium flow battery," *J Power Sources*, vol. 294, pp. 562–568, Oct. 2015, doi: 10.1016/J.JPOWSOUR.2015.06.123.

[26] Z. Jiang, K. Klyukin, K. Miller, and V. Alexandrov, "Mechanistic Theoretical Investigation of Self-Discharge Reactions in a Vanadium Redox Flow Battery," *J Phys Chem B*, vol. 123, no. 18, pp. 3976–3983, May 2019, doi: 10.1021/ACS.JPCB.8B10980.

[27] E. H. Kirk, F. Fenini, S. N. Oreiro, and A. Bentien, "Temperature-Induced Precipitation of $V_2O_5$ in Vanadium Flow Batteries—Revisited," *Batteries*, vol. 7, no. 4, p. 87, Dec. 2021, doi: 10.3390/BATTERIES7040087/S1.

[28] P. J. Alphonse and G. Elden, "The investigation of thermal behavior in a vanadium redox flow battery during charge and discharge processes," *J Energy Storage*, vol. 40, Aug. 2021, doi: 10.1016/j.est.2021.102770.

[29] "CRE 32-1 N-G-A-E-HQQE - 99392747 | Grundfos." Accessed: Oct. 14, 2025. [Online]. Available: https://product-selection.grundfos.com/hu/products/cr-cre-cri-crie-crn-crne-crt-crte/cre/cre-32-1-99392747?pumpsystemid=2788753113&tab=variant-curves

[30] J. Sun, Z. Guo, L. Pan, X. Fan, L. Wei, and T. Zhao, "Redox flow batteries and their stack-scale flow fields," Dec. 01, 2023, *Springer*. doi: 10.1007/s43979-023-00072-6.

[31] Z. Guo, J. Ren, J. Sun, B. Liu, X. Fan, and T. Zhao, "A bifurcate interdigitated flow field with high performance but significantly reduced pumping work for scale-up of redox flow batteries," *J Power Sources*, vol. 564, Apr. 2023, doi: 10.1016/j.jpowsour.2023.232757.

[32] P. A. Prieto-Díaz, A. A. Maurice, and M. Vera, "Measuring density and viscosity of vanadium electrolytes: A database with multivariate polynomial fits," *J Energy Storage*, vol. 94, p. 112429, Jul. 2024, doi: 10.1016/J.EST.2024.112429.

[33] "AEP Reflectivity-Emissivity", Accessed: Jul. 09, 2025. [Online]. Available: https://www.aepspan.com/files/AEP%20Reflectivity-Emissivity.pdf

[34] F. Fan, J. Wang, H. Pan, Z. Li, and D. Zhao, "Optical-thermal modeling and geographic analysis of dusty radiative cooling surfaces," *Renewable and Sustainable Energy Reviews*, vol. 205, p. 114878, Nov. 2024, doi: 10.1016/J.RSER.2024.114878.

[35] P. A. Prieto-Díaz, A. A. Maurice, M. Vera, G. de España, and E. Commission, "Measurements of density and viscosity of vanadium posolyte and negolyte," 2024, *e-cienciaDatos*. doi: doi/10.21950/M0OG5W.

[36] Z. Wei, J. Zhao, M. Skyllas-Kazacos, and B. Xiong, "Dynamic thermal-hydraulic modeling and stack flow pattern analysis for all-vanadium redox flow battery," *J Power Sources*, vol. 260, pp. 89–99, Aug. 2014, doi: 10.1016/j.jpowsour.2014.02.108.

[37] A. Tang, S. Ting, J. Bao, and M. Skyllas-Kazacos, "Thermal modelling and simulation of the all-vanadium redox flow battery," *J Power Sources*, vol. 203, pp. 165–176, Apr. 2012, doi: 10.1016/j.jpowsour.2011.11.079.

[38] D. W. Green and M. Z. Southard, Eds., *Perry's Chemical Engineers' Handbook*, 9th Edition. New York: McGraw-Hill Education, 2019. [Online]. Available: https://www.accessengineeringlibrary.com/content/book/9780071834087

[39] A. Patti and D. Acierno, "Thermal Conductivity of Polypropylene-Based Materials," in *Polypropylene*, W. Wang and Y. Zeng, Eds., Rijeka: IntechOpen, 2019, ch. 3. doi: 10.5772/intechopen.84477.

[40] L. Goddijn-Murphy and B. Williamson, "On thermal infrared remote sensing of plastic pollution in natural waters," *Remote Sens (Basel)*, vol. 11, no. 18, Sep. 2019, doi: 10.3390/RS11182159.

[41] *VDI Heat Atlas*. Berlin, Heidelberg: Springer Berlin Heidelberg, 2010. doi: 10.1007/978-3-540-77877-6.

[42] "Vanadium Flow Battery Energy Storage - Invinity." Accessed: Jul. 07, 2025. [Online]. Available: https://invinity.com/vanadium-flow-batteries/

[43] "Products – Cellcube." Accessed: Jul. 07, 2025. [Online]. Available: https://www.cellcube.com/products/

[44] "Product Variations | Vanadium Redox Flow Battery | Sumitomo Electric." Accessed: Jul. 07, 2025. [Online]. Available: https://sumitomoelectric.com/products/flow-batteries/product-variations

[45] A. Driemel *et al.*, "Baseline Surface Radiation Network (BSRN): Structure and data description (1992-2017)," *Earth Syst Sci Data*, vol. 10, no. 3, pp. 1491–1501, Aug. 2018, doi: 10.5194/ESSD-10-1491-2018.

[46] S. Bogdanov *et al.*, "Dynamic modeling of vanadium redox flow batteries: Practical approaches, their applications and limitations," *J Energy Storage*, vol. 57, p. 106191, Jan. 2023, doi: 10.1016/J.EST.2022.106191.

[47] X. L. Zhou, Y. K. Zeng, X. B. Zhu, L. Wei, and T. S. Zhao, "A high-performance dual-scale porous electrode for vanadium redox flow batteries," *J Power Sources*, vol. 325, pp. 329–336, Sep. 2016, doi: 10.1016/J.JPOWSOUR.2016.06.048.

[48] "TXBR-ECOWATT – S&P – S&P." Accessed: Jul. 31, 2025. [Online]. Available: https://www.solerpalau.com/en-en/cylindrical-cased-axial-flow-fans-txbr-ecowatt-1634-serie/

[49] K. B. Shyam Prasad, P. V. Suresh, and S. Jayanti, "A hydrodynamic network model for interdigitated flow fields," *Int J Hydrogen Energy*, vol. 34, no. 19, pp. 8289–8301, Oct. 2009, doi: 10.1016/j.ijhydene.2009.07.107.